\documentclass[showpacs,preprintnumbers,amsmath,amssymb,a4paper,superscriptaddress]{revtex4}
\usepackage{amsmath}
\usepackage{mathtools}
\usepackage{epsfig}
\usepackage{amssymb}
\usepackage{float}
\usepackage{graphicx, subfigure}
\usepackage{dcolumn}
\usepackage{bm}
\usepackage[font=small,labelfont=bf,labelsep=space]{caption}
\usepackage{refstyle}

\newcommand{\latinword}[1]{\textsf{\itshape #1}}
\def\nn{\nonumber}

\begin{document}
\title{Confinement-Induced Resonances in Two-Center Problem via Pseudopotential Approach}
\date{\today}
\author{Sara Shadmehri}
\email[]{shadmehri@theor.jinr.ru}\affiliation{Bogoliubov Laboratory of Theoretical Physics, Joint Institute for Nuclear Research, Dubna, Moscow Region 141980, Russian Federation}

\author{Vladimir S. Melezhik}
\email[]{melezhik@theor.jinr.ru} \affiliation{Bogoliubov Laboratory of Theoretical Physics, Joint Institute for Nuclear Research, Dubna, Moscow Region 141980, Russian Federation}
\affiliation{Peoples' Friendship University of Russia (RUDN University) Miklukho-Maklaya st. 6, Moscow,  117198, Russian Federation}%

\date{\today}

\begin{abstract}\label{txt:abstract}
We study confined scattering of a quantum particle by two centers fixed on the longitudinal axis of a harmonic waveguide-like trap. The conditions of confinement-induced resonances (CIRs) appearing in these systems, when scattering cross section approaches the unitary limit, are derived for a regularized pseudopotential describing particle interaction with scattering centers. In the limit of a single center, the position of CIR for even state tends to the well-known result obtained by Olshanii. Our result can be applicable to confined atomic scattering by fixed impurities, like ions or Rydberg atoms, with possible extension to $N$ impurities, or by two-atomic molecules.
\end{abstract}

\maketitle

\section{Introduction}
Experimental investigations of quantum gases stimulated the development of the theory of low-dimensional quantum systems in confined geometry of atomic traps \cite{Pethik,Pitaevskii,Leggett,Bloch,Chin,Dunjko}. So far, starting from the seminal works \cite{Busch,Olshanii}, the discrete as well as continuum spectra of the confined two-body problem have been investigated in detail (see, for example \cite{Bergeman,Bolda,Blume,Kim,Peano,Idziaszek2006,Kim2006,Melezhik2007,Kim2007} and  a review given in \cite{Dunjko}).
However, growing interest in cold hybrid atom-ion, atom-impurity and molecular systems \cite{Saffman,Schurer} makes it necessary to extend the theoretical analysis to the quantum two-center problem in confining traps. For example, a possibility is shown to model stationary nuclear cores and mobile electrons in a crystal with trapped ions and cold atoms  \cite{Bissbort}. Also, the use of trapped Rydberg atoms as impurities in place of ions for modeling other quantum processes is discussed \cite{Saffman}. The classical two-center problem (in free space) - a quantum particle in the field of two force centers situated at a fixed distance from each other - has a fundamental importance in molecular physics \cite{Slater,Komarov}. Its generalization to the case of an atom in the field of two impurities fixed inside a confining trap can have potentially interesting applications for studying more complicated confined atom-ion, atom-impurities and molecular systems.

The first step in the study of the discrete spectrum of the confined two-center problem was made in a recent paper \cite{Marta}. In the present work we study the continuum spectrum of the two-center problem confined in a harmonic waveguide trap (the continuum spectrum of an atom in the field of two impurities fixed on the longitudinal axis of the harmonic waveguide trap). A special consideration is devoted to finding the conditions of occurrence of the so-called confinement-induced resonances (CIRs)\cite{Olshanii} in the confined two-center problem. The CIRs were predicted by Olshanii in \cite{Olshanii} for the case of one scattering center in harmonic waveguide-like traps. The dynamics of this system was mapped to an effective 1D Hamiltonian with s-wave pseudopotential \cite{Fermi,Huang}, and the condition of appearance of CIR was formulated. It was shown that in the case of{\normalsize } CIR, the effective 1D coupling constant ($g_{1D}$) of the effective interaction was altered from strongly repulsive ($+\infty$) to attractive ($-\infty$) and the maximum, approaching the unitary limit, appeared in the reflection coefficient.

It was found that the theoretical description of the confined ultracold atomic systems and processes with contact interparticle interactions (pseudopotentials) gives a realistic picture here \cite{Leggett,Bloch,Chin,Dunjko}.
 The first proposed zero-range pseudopotential for s-wave scattering by Fermi-Huang \cite{Fermi,Huang} was extended for higher partial waves in \cite{Stock2005, Derevianko2005, Idziaszek2006, Stampfer2008} and non-zero energies \cite{Bolda,Olshani-Pri2001}. Different pseudopotentials were devised to tackle with various shapes or types of confinements; in \cite{Pricoupenko2007, Kanjilal2006} a 2D pseudopotential was developed for zero-range interaction in quasi-2D atomic gases and in \cite{Zhang2013} an alternative regularization method within pseudopotential was used to predict CIR in square-shape transversal confinement.

While the one-center problem in free space is spherically symmetric and s-wave pseudopotential is adequate here, the two-center problem does not possess this symmetry any more and demands special consideration. We perform the necessary modification of the pseudopotential approach which is shown to be applicable to confined scattering by two scattering centers and potentially can be extended to $N$-center problems as well.

In the following, we first discuss the necessary modification of the pseudopotential approach for the confined two-center problem (Sec.\ref{Alternative..}). In Sec.\ref{Continuous...}, the scattering amplitudes describing confined scattering by two-centers in a harmonic waveguide-like trap is derived in the pseudopotential approach. The reduction of the problem to an effective 1D Hamiltonian is given in Sec.\ref{1D-effective-zero-range} . Finally, we obtain the conditions of occurrence of CIRs in the confined two-center problem and discuss the dependence of the CIR positions on the parameters of the problem (Sec.\ref{Results}). In Sec.\ref{Conclusion}, a brief conclusion is given.

\section{Alternative Regularization Procedure for Pseudopotential}\label{Alternative..}
We study the scattering of a confined atom (of mass $m$) from two fixed centers (impurities) in a harmonic weveguide. The corresponding Hamiltonian is given by
\begin{eqnarray}\label{H3D}
H_{3D}= -\frac{\hbar^{2}}{2m}\frac{\partial^{2}}{\partial z^2 }+H_{\perp}+V_{3D}\,\,,
\end{eqnarray}
where
\begin{eqnarray}
H_{\perp}=-\frac{\hbar^{2}}{2m}(\frac{\partial^{2}}{\partial \rho^2 }+\frac{1}{\rho}\frac{\partial}{\partial\rho}+\frac{1}{\rho^2}\frac{\partial^2}{\partial \varphi^2})+\frac{1}{2}m\omega_{\perp}^2\rho^2
 \end{eqnarray}
is the Hamiltonian of a transversal 2D harmonic oscillator (with frequency $\omega_{\perp}$) describing interaction of the atom with confining trap, and $V_{3D}$ is the interaction potential between the atom and two impurities (see \figref{Fig1}).

When there is only single scatterer, the zero-range pseudopotential $V_{3D}$ can be presented as \cite{Olshanii}
\begin{eqnarray}\label{V3D-single}
V_{3D}\psi=g_{_{3D}}\delta^{3}(\textbf{r})\frac{\partial}{\partial r}(r\psi)\,\,,
\end{eqnarray}
where $g_{_{3D}}={2\pi\hbar^{2}a_{_{3D}}}/{m}$ is the coupling constant defined by the 3D scattering length $a_{_{3D}}$ in free space ($H_{\perp}=0$) \cite{Fermi, Huang}.

For the case of two impurities located at the points $z=a$ and $z=-a$ on the longitudinal Z-axis of the confining trap, one can assume $V_{3D}$ as in ref.\cite{Marta}
\begin{eqnarray}\label{wrong-V3D}
V_{_{3D}}\psi=\frac{1}{2}g_{_{3D}}\left[\delta^{3}(\textbf{r}_{1})\frac{\partial}{\partial r_1}(r_1\psi)+\delta^{3}(\textbf{r}_{2})\frac{\partial}{\partial r_2}(r_2\psi)\right] \,\,,
\end{eqnarray}
where $\textbf{r}_{1}=\textbf{r}-\textbf{\textit{a}}$ and $\textbf{r}_{2}=\textbf{r}+\textbf{\textit{a}}$ , $\textbf{\textit{a}}=a\textbf{n}_z$. Here we have multiplied their potential by factor $1/2$ so that it can mimic the single scatterer potential in the limit $a\rightarrow 0$. However, this potential is not capable of reproducing the Olshanii result \cite{Olshanii} when $a$ tends to zero, due to a especially chosen regularization in (\ref{wrong-V3D}) (see below and the text after Eq.(\ref{Olshanii fe})).

In search of a suitable regularization operator, we have to recall the enforcements of a zero-range interaction potential for the case of scattering from a single scatterer. According to \cite{Dunjko}, whenever the Hamiltonian contains such a potential, the wave function, $\psi(\textbf{r})$ , can be a solution of the corresponding Schr\"odinger equation in free space only if it obeys the Bethe-Peierls contact condition \cite{Bethe},
\begin{eqnarray}\label{original contact condition}
\psi(\textbf{r})=A(\frac{1}{r}-\frac{1}{a_{_{3D}}})+\mathcal{O}(r)~~~~as~r\rightarrow 0 \,\,.
\end{eqnarray}

Applying the regularization operator $\frac{\partial}{\partial r}(r.)$ in Eq.(\ref{V3D-single}) to the function (\ref{original contact condition}) results in
\begin{eqnarray}\label{regularization effect}
\frac{\partial}{\partial r}(r\psi)=-\frac{A}{a_{_{3D}}}\,\,,
\end{eqnarray}
i.e. it removes the divergence $A/r$ as $r\rightarrow 0$. However, a simple extension (\ref{wrong-V3D}) of the regularized pseudopotential (\ref{V3D-single}) to the case of two-center problem does not remove the singularity of the order $1/a$ in the scattering wave function $\psi({\bf{r}}) $ as $a\rightarrow 0$. Actually, the wave-function $\psi({\bf{r}}) $ has two singularities $A_1/r_1$ and $A_2/r_2$ as $r_1\rightarrow 0$, and $r_2\rightarrow 0$ \cite{Brueckner,Smirnov,Demkov,Baz,Ostrovskii},
\begin{eqnarray}
\psi({\bf{r}})=\frac{A_1}{r_1}+\frac{A_2}{r_2}+C \,\,,
\end{eqnarray}
then $V_{_{3D}}\psi$ in Eq.(\ref{wrong-V3D}) will be proportional to $\frac{1}{a}$, thus leading to a divergency as $a\rightarrow 0$.

To elimininate this drawback, we suggest alternative regularization $\frac{1}{2}\frac{\partial^2}{\partial r^2}(r^2.)$ instead of $\frac{\partial}{\partial r}(r.)$. It is clear that the action of the operator $\frac{1}{2}\frac{\partial^2}{\partial r^2}(r^2.)$ on the wave-function (\ref{original contact condition}) as $r\rightarrow 0$ is equivalent to (\ref{regularization effect}),
\begin{eqnarray}
\frac{1}{2}\frac{\partial^2}{\partial r^2}(r^2\psi)=-\frac{A}{a_{_{3D}}}\nn \,\,.
\end{eqnarray}

The extension of this procedure to the two-center problem removes the singularity $1/a$ as $a\rightarrow 0$ in $V_{3D}\psi$. We will see that the results obtained in the limit $a=0$ are in agreement with the one-center known counterparts (see Eq.(\ref{fe_simplified},\ref{fo_simplified}) and the text followed).

So using an alternative regularization procedure, we define the pseudopotential $V_{3D}$ modeling atom-impurities interaction as
\begin{eqnarray}\label{V_3D_elaborated}
V_{3D}\psi=\frac{1}{2}\times\frac{1}{2}g_{_{3D}}\left[\delta^3({\bf{r_1}})\frac{\partial^{2}}{\partial{r_1}\partial{r_2}}(r_1 r_2\psi)+\delta^3({\bf{r_2}})\frac{\partial^2}{\partial{r_2}\partial{r_1}}(r_2r_1\psi) \right] \,\,.
\end{eqnarray}

With the same justification, we arrive at the fact that the regularizing operator for the case of N impurities can be defined as $\frac{1}{N}\times\frac{1}{N!}\frac{\partial^N}{\partial r_1\partial r_2...\partial r_N}(r_1r_2...r_N.)$.

\section{Continuous spectrum of confined two-center problem in pseudopotential approach} \label{Continuous...}
Our goal is to find the scattering states of the Hamiltonian (\ref{H3D})  with the interaction potential defined in (\ref{V_3D_elaborated}). Assuming the wave function $\Psi({\b{r}})$ as expansion over eigenstates of the tranverse Hamiltonian $H_{\perp}$, i.e. $\phi_{n}(\rho)$, we have
\begin{eqnarray}\label{psi_main-expansion}
\Psi({\bf{r}})=\psi(z,\rho)=\sum_{n=0}^{\infty}\psi_{n}(z)\phi_{n}(\rho) \,\,.
\end{eqnarray}
Since at the position of scattering centers, only the transverse eigenstate with $m=0$ has a nonzero value at $\rho=0$, we have neglected the azimuthal angular dependence of the wave function and considered the transversal eigenstates as $\phi_{n}(\rho)=\langle \rho ,\varphi |n,m=0\rangle$.
By inserting the above wave function into the Schr\"odinger equation, we reach
\begin{eqnarray}\label{preliminary schrodinger eq}
\sum_{n=0}^{\infty}\left[-\frac{\hbar^{2}}{2m}\frac{\partial^{2}\psi_{n}}{\partial z^2 }\phi_{n}(\rho)+\psi_{n}(z)H_{\perp}\phi_{n}(\rho) \right]+V_{3D}\psi(z,\rho)=E\psi(z,\rho) \,\,.
\end{eqnarray}

With regard to the defined interaction potential (\ref{V_3D_elaborated}), $\psi_n(z)$ can be a solution of Eqs.(\ref{preliminary schrodinger eq}) only if some delta functions appear in its second derivative as follows (for some $\mu_{n_{\pm}} $ ):
\begin{eqnarray}\label{ansatz_second derivative}
\frac{\partial^{2}\psi_{n}}{\partial z^2 }=\latinword{R}\psi^{''}_{n}(z)+\mu_{n_{-}}\delta(z-a)+\mu_{n_{+}}\delta(z+a) \,\,,
\end{eqnarray}
where $\latinword{R}\psi^{''}_{n}(z)$ is the smooth continuous part of $\frac{\partial^{2}\psi_{n}}{\partial z^2 }$ with finite values at the points $z=\pm a$ and $\latinword{R}\psi^{''}_{n}(z)=\frac{d^{2}\psi_{n}}{dz^2 }$ at $z\neq\pm a$. Then, the terms containing delta functions
\begin{eqnarray}\label{main omega definition}
\Omega=\sum_{n=0}^{\infty}\left(\left[-\frac{\hbar^{2}}{2m}\mu_{n_{-}}\delta(z-a)-\frac{\hbar^{2}}{2m}\mu_{n_{+}}\delta(z+a)\right]\phi_{n}(\rho)\right)+V_{3D}\psi(z,\rho)
\end{eqnarray}
should be canceled among themselves, resulting in $\Omega =0 $ (the jump of the kinetic energy at the points $z\pm a$ is canceled by the contact interaction in these points).

We integrate both sides of Eq.(\ref{preliminary schrodinger eq}) over $\rho$ with the weight factor $2\pi\rho\phi_n^{\ast}(\rho)$; considering $\Omega =0 $ and the fact that away from $z=\pm a$, $\latinword{R}\psi^{''}_{n}(z)=\frac{d^{2}\psi_{n}}{dz^2 }$ , we get uncoupled ordinary differential equations
\begin{eqnarray}\label{1d schrodinger eq}
-\frac{\hbar^{2}}{2m}\frac{d^{2}\psi_{n}}{dz^2 }+(2n+1)\hbar\omega_{\perp}\psi_{n}(z)=E\psi_{n}(z)~~,~~~z\neq\pm a \,\,.
\end{eqnarray}

\begin{figure}[H]
\vskip -0.1in
\centering\includegraphics[width=0.8\textwidth]{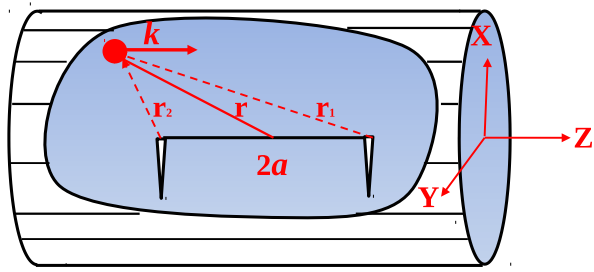}
\vskip -1.5in
\caption{A schematic representation of an atom with the wave number $k$ confined in a waveguide and scattered from two impurities simulated by a double-delta function potential }
\label{fig:Fig1}
\end{figure}

Thus, for $n=0$, we have
\begin{eqnarray}
\frac{d^2\psi_0}{dz^2}+k^2\psi_0=0~~,~~ k^2=\frac{2m}{\hbar^2}(E-\hbar\omega_\perp)>0 \,\,,
\end{eqnarray}
whose solution would be a superposition of $e^{ikz} $ and $ e^{-ikz}$. For an atom coming from $z=-\infty$, reflected and transmitted from pseudopotentials located at $z=-a$ and $z=a$ (\figref{Fig1}), the wave function, $\psi_0$, can be written as
\begin{eqnarray}
\psi_0(z)=\left\{\begin{array}{c}e^{ikz}+f^-e^{-ikz}~~,~~ z<-a \nn\\
A e^{ikz} + B e^{-ikz}~~,~~ -a<z<a \nn\\
e^{ikz}+f^+e^{ikz}~~,~~ z>a\end{array}\right.,\\
\end{eqnarray}
where $f^{\pm}(k)$ denote the atom-impurity forward and backward scattering amplitudes in the presence of the external confining potential. The forward-backward amplitudes $f^{\pm}(k)$ can be written as a sum $f^{\pm}(k)=f_e(k)\pm f_o(k)$ of even-odd (gerade-ungerade) scattering amplitudes $f_e(k)$ and $f_o(k)$.

Considering the asymptotic wave function as
\begin{eqnarray}\label{asymptotic-psi0}
\psi_0(z\rightarrow\pm\infty)=e^{ikz}+\left[f_e+\emph{sgn}(z)f_o\right]e^{ik\left|z\right|}
\end{eqnarray}
and applying continuity of the wave function at $z=a$ and $z=-a$ leads us to
\begin{eqnarray}\label{psi_0}
\psi_0(z)=e^{ikz}+A_0e^{ik\left| z-a\right|}+B_0e^{ik\left| z+a\right|} \,\,,
\end{eqnarray}
where
\begin{eqnarray}\label{A&B}
\left\{\begin{array}{c}A_0=\frac{f_e}{2\cos(ka)}+i\frac{f_o}{2\sin(ka)} \\ \\
B_0=\frac{f_e}{2\cos(ka)}-i\frac{f_o}{2\sin(ka)} \end{array}\right.
\end{eqnarray}

For $n\geqslant 1$, (\ref{1d schrodinger eq}) follows
\begin{eqnarray}
\frac{d^2\psi_{n}}{dz^2}-k_{n}^2\psi_{n}=0 \,\,,
\end{eqnarray}
where $k_{n}^2=\frac{2m}{\hbar^2}\left[(2n+1)\hbar\omega_\perp -E\right]$ and since $E=\hbar\omega_\perp +{\hbar^2k^2}/{(2m)}$, we have $k_{n}^2=\frac{2m}{\hbar^2}\left( 2n\hbar\omega_\perp - \frac{\hbar^{2}k^2}{2m}\right)$. Considering $a_{\perp}=\sqrt{\frac{\hbar}{m\omega_{\perp}}}$ and $\epsilon=-({\frac{a_{\perp}k}{2})}^2$ , we have
\begin{eqnarray}
k_n=\frac{2}{a_{\perp}}\sqrt{n+\epsilon} \,\,.
\end{eqnarray}

The asymptotic boundary condition for $\psi_n(z)$ is
\begin{eqnarray}
\psi_n(z\rightarrow\pm\infty)=0 \,\,.
\end{eqnarray}

Considering the above condition, we have the even and odd wave functions as follows:
\begin{eqnarray}
\psi_{n,e}(z)=\left\{\begin{array}{c}C_ne^{k_nz}~~,~~ z<-a \\
A_n\left(e^{k_nz}+e^{-k_nz}\right)~~,~~ -a<z<a \\
C_ne^{-k_nz}~~,~~ z>a\end{array}\right.
\end{eqnarray}
\begin{eqnarray}
\psi_{n,o}(z)=\left\{\begin{array}{c}-D_ne^{k_nz}~~,~~ z<-a \\
B_n\left(e^{k_nz}-e^{-k_nz}\right)~~,~~ -a<z<a \\
D_ne^{-k_nz}~~,~~ z>a\end{array}\right.
\end{eqnarray}

After applying continuity of the wave function at $z=a$ and $z=-a$, we obtain
\begin{eqnarray}
\psi_{n,e}(z)&=&\frac{C_n}{2\cosh(k_na)}\left[e^{-k_n|z-a|}+e^{-k_n|z+a|}\right]\label{psi_n_even}\\
\psi_{n,o}(z)&=&\frac{D_n}{2\sinh(k_na)}\left[e^{-k_n|z-a|}-e^{-k_n|z+a|}\right]\label{psi_n_odd} \,\,.
\end{eqnarray}

Thus, by using (\ref{psi_0}), (\ref{psi_n_even}), (\ref{psi_n_odd}), and considering $\psi_n$ as a superposition of even and odd wave functions ($\psi_n=\psi_{n,e}+\psi_{n,o}$), we have
\begin{eqnarray}
-\frac{\hbar^{2}}{2m}\frac{d^{2}\psi_{0}}{dz^2 }&=&\frac{\hbar^2k^2}{2m}\psi_0 -i\frac{\hbar^2k}{m}\left[A_0\delta(z-a)+B_0\delta(z+a)\right] \label{d2psi_0}\\ \nn\\
 -\frac{\hbar^{2}}{2m}\frac{d^{2}\psi_{n}}{dz^2 }&=&-\frac{\hbar^2k_n^2}{2m}\psi_n \nn\\&+& \frac{\hbar^2k_nC_n}{2m\cosh(k_na)}\left[\delta(z-a)+\delta(z+a)\right] \nn\\&+&\frac{\hbar^2k_nD_n}{2m\sinh(k_na)}\left[\delta(z-a)-\delta(z+a) \right]~,~~~n\geqslant 1 \,\,, \label{d2psi_n}
\end{eqnarray}
which is consistent with our assumption in Eq.(\ref{ansatz_second derivative}).

Now we have to clarify the last term in Eq.(\ref{main omega definition}), $V_{3D}\psi(z,\rho)$. By defining the coefficients $\eta_1$ and $\eta_2$ as
\begin{eqnarray}
\eta_1=\frac{1}{2}\frac{\partial^{2}}{\partial{r_1}\partial{r_2}}(r_1 r_2\psi)|_{r_1\rightarrow 0}\\ \nn \\
\eta_2=\frac{1}{2}\frac{\partial^{2}}{\partial{r_2}\partial{r_1}}(r_2 r_1\psi)|_{r_2\rightarrow 0} \,\,,
\end{eqnarray}
which turn out to be (see Appendix A)
\begin{eqnarray}\label{eta_1-definition}
\eta_1=\frac{1}{2}\frac{d^{2}}{dz^2}\left[(z-a)(z+a)\psi(\rho=0,z) \right]_{z\rightarrow a^+}
\end{eqnarray}
\begin{eqnarray}\label{eta_2-definition}
\eta_2=\frac{1}{2}\frac{d^2}{dz^2}\left[(z-a)(z+a)\psi(\rho=0,z) \right]_{z\rightarrow {-a}^-} \,\,,
\end{eqnarray}
and taking into account $\delta^3(\textbf{r})=\delta(\rho)\delta(z)/(2\pi\rho)$, we can represent $V_{3D}\psi$ in  Eq.(\ref{V_3D_elaborated}) as the following:
\begin{eqnarray}\label{V_3D_symlified}
V_{3D}\psi=\frac{1}{2}\frac{g_{_{3D}}\delta(\rho)}{2\pi\rho}\left[\eta_1\delta(z-a)+\eta_2\delta(z+a) \right] \,\,.
\end{eqnarray}

Now by inserting the irregular part of Eq.(\ref{d2psi_0}) and Eq.(\ref{d2psi_n}) (containing delta functions) and $V_{3D}$ as Eq.(\ref{V_3D_symlified}) into Eq.(\ref{main omega definition}), we come to the following relation:
\begin{eqnarray}
\Omega=-i\frac{\hbar^2k}{m}\left[A_0\delta(z-a)+B_0\delta(z+a) \right]\phi_0(\rho)&+&\frac{\hbar^2}{2m}\sum_{n=1}^{\infty}\left[\frac{C_nk_n}{\cosh(k_na)}\left[\delta(z-a)+\delta(z+a) \right]\phi_n(\rho) \right] \nn\\&+&\frac{\hbar^2}{2m}\sum_{n=1}^{\infty}\left[\frac{D_nk_n}{\sinh(k_na)}\left[\delta(z-a)-\delta(z+a) \right]\phi_n(\rho) \right]\nn\\&+&\frac{g_{_{3D}}\delta(\rho)}{4\pi\rho}\left[\eta_1\delta(z-a)+\eta_2\delta(z+a) \right] \,\,.
\end{eqnarray}

The demand $\Omega=0$ leads to two relations at $z=a$ and $z=-a$:
\begin{eqnarray}\label{Omega=0_z=a}
-i\frac{\hbar^2k}{m}A_0\phi_0(\rho)+\frac{\hbar^2}{2m}\sum_{n=1}^{\infty}\left[\left(\frac{C_n}{\cosh(k_na)}+\frac{D_n}{\sinh(k_na)} \right)k_n\phi_n(\rho) \right]+\frac{g_{_{3D}}}{4\pi\rho}\delta(\rho)\eta_1=0
\end{eqnarray}
\begin{eqnarray}\label{Omega=0_z=-a}
-i\frac{\hbar^2k}{m}B_0\phi_0(\rho)+\frac{\hbar^2}{2m}\sum_{n=1}^{\infty}\left[\left(\frac{C_n}{\cosh(k_na)}-\frac{D_n}{\sinh(k_na)} \right)k_n\phi_n(\rho) \right]+\frac{g_{_{3D}}}{4\pi\rho}\delta(\rho)\eta_2=0 \,\,.
\end{eqnarray}

By integrating Eq.(\ref{Omega=0_z=a}) and Eq.(\ref{Omega=0_z=-a}) over $\rho$ with the weight factor $2\pi\rho\phi_0^{\ast}(\rho)$ and taking into account $\phi_0^{\ast}(\rho=0)=1/{(a_\perp\sqrt{\pi})}$, we find the coefficients $A_0$ and $B_0$ in terms of $\eta_1$ and $\eta_2$:
\begin{eqnarray}
A_0=-i\frac{\sqrt{\pi}a_{_{3D}}}{ka_{\perp}}\eta_1\\
B_0=-i\frac{\sqrt{\pi}a_{_{3D}}}{ka_{\perp}}\eta_2 \,\,.
\end{eqnarray}

By using the definitions (\ref{A&B}), one can represent the scattering amplitudes $f_e$ and $f_o$ in (\ref{asymptotic-psi0}) as
\begin{eqnarray}
f_e&=&-i\frac{\sqrt{\pi}a_{_{3D}}}{ka_{\perp}}\cos(ka)(\eta_1+\eta_2)\label{f_even-main definition}\\
f_o&=&-\frac{\sqrt{\pi}a_{_{3D}}}{ka_{\perp}}\sin(ka)(\eta_1-\eta_2)\label{f_odd-main definition} \,\,.
\end{eqnarray}

Likewise, by integrating Eq.(\ref{Omega=0_z=a}) and Eq.(\ref{Omega=0_z=-a}) over $\rho$ with the weight factor $2\pi\rho\phi_n^{\ast}(\rho)$ and taking into account $\phi_n^{\ast}(\rho=0)=1/{(a_\perp\sqrt{\pi})}$, we find the coefficients $C_n$ and $D_n$
\begin{eqnarray}
C_n=-\frac{\sqrt{\pi}a_{_{3D}}}{k_na_{\perp}}\cosh(k_na)(\eta_1+\eta_2)\\
D_n=-\frac{\sqrt{\pi}a_{_{3D}}}{k_na_{\perp}}\sinh(k_na)(\eta_1-\eta_2) \,\,.
\end{eqnarray}

By inserting the found coefficients $A_0$, $B_0$, $C_n$ and $D_n$ into Eq.(\ref{psi_0}), Eq.(\ref{psi_n_even}) and Eq.(\ref{psi_n_odd}) and, finally, in the expansion Eq.(\ref{psi_main-expansion}), we reach to the following relation for the wave function $\psi(z,\rho)$
\begin{eqnarray}
\psi(z,\rho)=e^{ikz}\phi_0(\rho)-i\frac{\sqrt{\pi}a_{_{3D}}}{ka_{\perp}}\eta_1e^{ik|z-a|}\phi_0(\rho)-i\frac{\sqrt{\pi}a_{_{3D}}}{ka_{\perp}}\eta_2e^{ik|z+a|}\phi_0(\rho)\nn\\-\sum_{n=1}^{\infty}\left[\frac{\sqrt{\pi}a_{_{3D}}}{k_na_{\perp}}\eta_1\phi_n(\rho)e^{-k_n|z-a|}+\frac{\sqrt{\pi}a_{_{3D}}}{k_na_{\perp}}\eta_2\phi_n(\rho)e^{-k_n|z+a|} \right] \,\,.
\end{eqnarray}

At $\rho=0$, it reduces to
\begin{eqnarray}\label{initial_psi-rho=0}
\psi(z,\rho=0)=\frac{e^{ikz}}{a_{\perp}\sqrt{\pi}}&-&i\frac{a_{_{3D}}}{ka_\perp^2}\left[\eta_1e^{ik|z-a|}+\eta_2e^{ik|z+a|} \right]\nn\\&-& \frac{a_{_{3D}}}{2a_\perp}\left[\eta_1\sum_{n=1}^{\infty}\frac{e^{-\frac{2|z-a|\sqrt{n+\epsilon}}{a_\perp}}}{\sqrt{n+\epsilon}}+\eta_2\sum_{n=1}^{\infty}\frac{e^{-\frac{2|z+a|\sqrt{n+\epsilon}}{a_\perp}}}{\sqrt{n+\epsilon}} \right] \,\,.
\end{eqnarray}

Following the computational scheme suggested in \cite{Olshanii} (see also \cite{Dunjko}) for the summation
\begin{eqnarray}
\Lambda(x,\epsilon)=\sum_{n=1}^{\infty}\frac{e^{-x\sqrt{n+\epsilon}}}{\sqrt{n+\epsilon}}
\end{eqnarray}
appearing in Eq.(\ref{initial_psi-rho=0}), we isolate the divergencies at $z=\pm a$ by adding to and subtracting $2/x$ from the above expression
\begin{eqnarray}\label{Lambda_main definition}
\Lambda(x,\epsilon)=\frac{2}{x}+\widetilde{\Lambda}(x,\epsilon) \,\,.
\end{eqnarray}

Then, the wave function at $\rho=0$ can be reduced to the form
\begin{eqnarray}\label{main psi_z}
\psi(z,\rho=0)&=&-\frac{a_{_{3D}}\eta_1}{2|z-a|}-\frac{a_{_{3D}}\eta_2}{2|z+a|}\nn\\&+&\frac{e^{ikz}}{a_{\perp}\sqrt{\pi}}-i\frac{a_{_{3D}}}{ka_\perp^2}\left[\eta_1e^{ik|z-a|}+\eta_2e^{ik|z+a|} \right]\nn\\&-& \frac{a_{_{3D}}}{2a_\perp}\left[\eta_1 \widetilde{\Lambda}(\frac{2|z-a|}{a_{\perp}},\epsilon)+\eta_2 \widetilde{\Lambda}(\frac{2|z+a|}{a_{\perp}},\epsilon) \right] \,\,.
\end{eqnarray}

To obtain the scattering amplitudes (Eqs.(\ref{f_even-main definition}, \ref{f_odd-main definition})), we need to find an expression for $\eta_1$ and $\eta_2$. According to the definitions (\ref{eta_1-definition}) and (\ref{eta_2-definition}), it is necessary to find the second derivative of the wave function (\ref{main psi_z}) multiplied by the factor $(z-a)(z+a)$ in the respective limits ($z\rightarrow \pm a$).

It is clear that after calculating $\frac{d^{2}}{dz^2}\left[(z-a)(z+a)\psi(z,\rho=0) \right]$ with $\psi$ defined in Eq.(\ref{main psi_z}), the first two terms in Eq.(\ref{main psi_z}) (diverging at $z=\pm a$) are removed, as it is desired. So only the remaining terms in Eq.(\ref{main psi_z}) matter, which we define as $S(z)$ in $\psi(z,\rho=0)=-\frac{a_{_{3D}}\eta_1}{2|z-a|}-\frac{a_{_{3D}}\eta_2}{2|z+a|}+S(z)$. Then we have
\begin{eqnarray}\label{2-nd derivative}
\frac{1}{2}\frac{d^{2}}{dz^2}\left[(z-a)(z+a)\psi(z,\rho=0) \right]=S(z)+2z\frac{dS}{dz}+\frac{1}{2}(z^2-a^2)\frac{d^2S}{dz^2}~,
\end{eqnarray}
a finite function at $z=\pm a$. Here, we need to find the derivative of the sum $\widetilde{\Lambda}(x,\epsilon)$ over $z$. By using its definition (Eq.(\ref{Lambda_main definition})), we obtain
\begin{eqnarray}
\frac{d\widetilde{\Lambda}}{dz}&=&\frac{d\widetilde{\Lambda}}{dx}\frac{dx}{dz}\nn\\&=&\left[\frac{2}{x^2} -\sum_{n=1}^{\infty}e^{-x\sqrt{n+\epsilon}}  \right]\frac{dx}{dz} \,\,,
\end{eqnarray}
where $\frac{dx}{dz}=\frac{2}{a_{\perp}}\frac{z\pm a}{|z\pm a|} $. Defining the summation appearing here as
\begin{eqnarray}\label{Fbar_definition}
\sum_{n=1}^{\infty}e^{-x\sqrt{n+\epsilon}}=\frac{2}{x^2}+\widetilde{F}(x,\epsilon)~,
\end{eqnarray}
we conclude
\begin{equation}
\frac{\partial\widetilde{\Lambda}}{\partial z}=-\widetilde{F}(\frac{2}{a_{\perp}}|z\pm a|,\epsilon)\frac{2}{a_{\perp}}\frac{z\pm a}{|z\pm a|} \,\,.
\end{equation}

We need to find an expresion for $\widetilde{F}(x,\epsilon)$, when $x$ goes to zero. Following the scheme suggested in \cite{Dunjko} for computing $\widetilde{\Lambda}(x,\epsilon)$, we represent $\widetilde{F}(x,\epsilon)$ as follows:
\begin{eqnarray}
\widetilde{F}(x,\epsilon)=-\frac{2}{x^2}+\int_{1}^{\infty}e^{-x\sqrt{u+\epsilon}}du-\lim_{N\rightarrow\infty}\int_{1}^{N}e^{-x\sqrt{u+\epsilon}}du+\lim_{N\rightarrow\infty}\sum_{n=1}^{N}e^{-x\sqrt{n+\epsilon}}
\end{eqnarray}
and reduce it to the form
\begin{eqnarray}
\widetilde{F}(x,\epsilon)&=&-\frac{2}{x^2}+\frac{2}{x^2}(1+x\sqrt{1+\epsilon})e^{-x\sqrt{1+\epsilon}}\nn\\
&+&\lim_{N\rightarrow\infty}\left[2\frac{(1+x\sqrt{N+\epsilon})e^{-x\sqrt{N+\epsilon}}-(1+x\sqrt{1+\epsilon})e^{-x\sqrt{1+\epsilon}}}{x^2}+\sum_{n=1}^{N}e^{-x\sqrt{n+\epsilon}}    \right] \,\,,
\end{eqnarray}
which is convergent as $x\rightarrow 0$
\begin{eqnarray}
\widetilde{F}(0,\epsilon)= -\epsilon \,\,.
\end{eqnarray}

In order to find an expression for $\widetilde{\Lambda}(0,\epsilon)$, we follow the same procedure and finally obtain
\begin{eqnarray}
\widetilde{\Lambda}(0,\epsilon)=\lim\limits_{N\to\infty}\left[-2\sqrt{N+\epsilon}+\sum_{n=1}^{N}\frac{1}{\sqrt{n+\epsilon}} \right]=\zeta(1/2,1+\epsilon) \,\,,
\end{eqnarray}
where $\zeta(1/2,x)$ represents the Hurwitz zeta function which is known as
\begin{eqnarray}
\zeta(1/2,x)=\lim\limits_{N\to\infty}\left[-2\sqrt[\uparrow]{N+x}+\sum_{n=0}^{N}\frac{1}{\sqrt[\downarrow]{n+\epsilon}} \right] \,\,,
\end{eqnarray}
where $\sqrt[\uparrow]{|r|e^{i\theta}}=\sqrt{|r|}e^{i\theta/2}$ for $0\leqslant\theta<2\pi $ and $\sqrt[\downarrow]{|r|e^{i\theta}}=\sqrt{|r|}e^{i\theta/2}$ for $-2\pi<\theta\leqslant0 $.

Now using (\ref{eta_1-definition}, \ref{eta_2-definition}) and (\ref{2-nd derivative}), we reach two coupled equations with respect to unknown $\eta_1$ and $\eta_2$,
\begin{eqnarray}
\eta_1 &=&\frac{e^{ika}}{a_\perp\sqrt{\pi}}-i\frac{a_{_{3D}}}{ka_\perp^2}\left(\eta_1+\eta_2 e^{i2ka} \right)-\frac{a_{_{3D}}}{2a_\perp}\left[\eta_1\widetilde{\Lambda}(0,\epsilon)+\eta_2\widetilde{\Lambda}(\frac{4a}{a_\perp},\epsilon) \right]\nn\\ &+& \frac{2aike^{ika}}{a_{\perp}\sqrt{\pi}}+\frac{2aa_{_{3D}}}{a_{\perp}^2}(\eta_1+\eta_2e^{2ika})-\frac{2aa_{_{3D}}}{a_{\perp}^2}\left[\epsilon\eta_1-\widetilde{F}(\frac{4a}{a_{\perp}},\epsilon)\eta_2 \right]
\end{eqnarray}
\begin{eqnarray}
\eta_2 &=&\frac{e^{-ika}}{a_\perp\sqrt{\pi}}-i\frac{a_{_{3D}}}{ka_\perp^2}\left(\eta_1 e^{i2ka}+\eta_2  \right)-\frac{a_{_{3D}}}{2a_\perp}\left[\eta_1\widetilde{\Lambda}(\frac{4a}{a_\perp},\epsilon)+\eta_2\widetilde{\Lambda}(0,\epsilon) \right]\nn\\ &-& \frac{2aike^{-ika}}{a_{\perp}\sqrt{\pi}}+\frac{2aa_{_{3D}}}{a_{\perp}^2}(\eta_1e^{2ika}+\eta_2)-\frac{2aa_{_{3D}}}{a_{\perp}^2}\left[\epsilon\eta_2-\widetilde{F}(\frac{4a}{a_{\perp}},\epsilon)\eta_1 \right] \,\,,
\end{eqnarray}
which can be represented as
\begin{eqnarray}\label{coupled equation}
\left\{\begin{array}{c}G\eta_1+H\eta_2 = \frac{e^{ika}}{a_{\perp}\sqrt{\pi}}(1+2iak)=\chi\\ \\H\eta_1+G\eta_2 =\frac{e^{-ika}}{a_{\perp}\sqrt{\pi}}(1-2iak)=\chi^{\ast} \end{array} \right.
\end{eqnarray}
where
\begin{eqnarray}
G&=&1+\frac{a_{_{3D}}}{2a_\perp}\widetilde{\Lambda}(0,\epsilon)+i\frac{a_{_{3D}}}{ka_\perp^2}-\frac{2aa_{_{3D}}}{a_{\perp}^2}(1-\epsilon)\\H&=&\frac{a_{_{3D}}}{2a_\perp}\widetilde{\Lambda}(\frac{4a}{a_\perp},\epsilon)+i\frac{a_{_{3D}}}{ka_\perp^2}e^{2ika}-\frac{2aa_{_{3D}}}{a_{\perp}^2}(e^{2ika}+\widetilde{F}(\frac{4a}{a_{\perp}},\epsilon)) \,\,.
\end{eqnarray}

By solving Eq.(\ref{coupled equation}), we reach the expression for $\eta_1$ and $\eta_2$
\begin{eqnarray}
\eta_1&=&\frac{ G\chi -H{\chi}^{\ast}}{G^2-H^2}\\
\eta_2&=& \frac{G{\chi}^{\ast}-H\chi}{G^2-H^2} \,\,.
\end{eqnarray}

By inserting the above values into the definition of scattering amplitudes (Eq.(\ref{f_even-main definition}) and Eq.(\ref{f_odd-main definition})), we obtain the scattering amplitudes as
\begin{eqnarray}\label{fe_simplified}
f_e=-\frac{\cos^2(ka)-ka\sin(2ka)}{\frac{1}{2}(1+e^{2ika})-ika_{\perp}\left[\frac{a_\perp}{2a_{_{3D}}}+\frac{1}{4}\left(\widetilde{\Lambda}(0,\epsilon)+\widetilde{\Lambda}(\frac{4a}{a_\perp},\epsilon)\right)-\frac{a}{a_{\perp}}\left(1-\epsilon +e^{2ika}+\widetilde{F}(\frac{4a}{a_{\perp}},\epsilon)\right)\right]}\nn \\
\end{eqnarray}
\begin{eqnarray}\label{fo_simplified}
f_o=-\frac{\sin^2(ka)+ka\sin(2ka)}{\frac{1}{2}(1-e^{2ika})-ika_{\perp}\left[\frac{a_\perp}{2a_{_{3D}}}+\frac{1}{4}\left(\widetilde{\Lambda}(0,\epsilon)-\widetilde{\Lambda}(\frac{4a}{a_\perp},\epsilon)\right)-\frac{a}{a_{\perp}}\left(1-\epsilon -e^{2ika}-\widetilde{F}(\frac{4a}{a_{\perp}},\epsilon)\right)\right]} \nn \,\,.\\
\end{eqnarray}

It is clear that in the limit $a\to 0$, the even scattering amplitude corresponds to the value obtained by Olshanii for the confined scattering on a single center in the pseuopotential approach \cite{Olshanii, Dunjko}
\begin{eqnarray}\label{Olshanii fe}
f_e=-\frac{i}{{ka_\perp}/{2}}\frac{1}{\frac{a_\perp}{a_{_{3D}}}+\frac{2i}{ka_\perp}+\widetilde{\Lambda}(0,\epsilon)}
\,\,.
\end{eqnarray}

Using the fact that $\widetilde{\Lambda}(0,\epsilon)=\zeta({1}/{2},1+\epsilon)=\zeta({1}/{2},\epsilon)-{1}/{\sqrt[\downarrow]{\epsilon}}$ and $\epsilon=-{a_\perp^2k^2}/4$ (${1}/{\sqrt[\downarrow]{\epsilon}}=2i/ka_{\perp}$) leads us to Eq.(65) of ref.\cite{Dunjko} for the even scattering amplitude. We have to note that using the interaction potential in the form of Eq.(\ref{wrong-V3D}) does not remove the diverging terms ,$\frac{1}{|z-a|}$ and $\frac{1}{|z+a|}$, of wave function (\ref{main psi_z}) and leads us to a divergency in the final result. Actually, with potential (\ref{wrong-V3D}) and in the limit $a\rightarrow 0$, the singular term $+\frac{a_{\perp}}{4a}$ appears in the denominator of Eq.(\ref{Olshanii fe}), obviously deviating from the Olshanii result.

The odd scattering amplitude $f_o$ defined by Eq.(\ref{fo_simplified}) tends to zero in the limit of a single s-wave zero-range potential as $a\rightarrow 0$.

\section{1D effective zero-range potential} \label{1D-effective-zero-range}
In the papers \cite{Olshanii, Dunjko} an effective 1D theory was suggested for approximating CIRs. This approach turned out to be very convenient and efficient in experimental analysis of CIRs \cite{Haller, Zurn}. Following the idea of \cite{Olshanii, Dunjko}, we eliminate the confined degree of freedom from the 3D Hamiltonian and reduce the problem to an effective 1D Hamiltonian with a 1D effective interaction potential. By rewriting the delta functions in Eq.(\ref{d2psi_0}) as $-i\frac{\hbar^2k}{m}A_0\frac{\psi_0(z)}{\psi_0(z=a)}\delta(z-a)$ and $-i\frac{\hbar^2k}{m}B_0\frac{\psi_0(z)}{\psi_0(z=-a)}\delta(z+a)$, we get the effective 1D Schr\"odinger equation
\begin{eqnarray}\label{effective-1D-Schrodinger eq}
-\frac{\hbar^{2}}{2m}\frac{d^{2}\psi_{0}}{dz^2 }+V_{1D}\psi_0(z)=\frac{\hbar^2k^2}{2m}\psi_0(z)\,\,,
\end{eqnarray}
where the 1D effective potential is expressed as
\begin{eqnarray}\label{V1D}
V_{1D}=\frac{1}{2}\left[g^{+}_{_{1D}}\delta(z-a)+g^{-}_{_{1D}}\delta(z+a)\right]\,\,,
\end{eqnarray}
and by using the representation (\ref{psi_0}) for the wave function $\psi_0(z)$, we come to the following coupling constants:
\begin{eqnarray}\label{g1D definition}
g^{+}_{_{1D}}=i\frac{k\hbar^2}{m}e^{-ika}\frac{{f_e}/\cos(ka)+i{f_o}/\sin(ka)}{1+f_e+f_o}
\end{eqnarray}
\begin{eqnarray}\label{g1D definition-2}
g^{-}_{_{1D}}=i\frac{k\hbar^2}{m}e^{-ika}\frac{{f_e}/\cos(ka)-i{f_o}/\sin(ka)}{e^{-2ika}+f_e-f_o} \,\,.
\end{eqnarray}

These constants can be alternatively derived from the jump condition imposed on the first derivative of $\psi_0(z)$ due to the appearance of delta functions in the effective potential (\ref{V1D})
\begin{eqnarray}\label{jump-condition-1}
\left(\frac{d\psi_0}{dz}\right)_{z=a^+}-\left(\frac{d\psi_0}{dz}\right)_{z=a^-}=\frac{2m}{\hbar^2}g^{+}_{_{1D}}\psi_0(z=a)
\end{eqnarray}
\begin{eqnarray}\label{jump-condition-2}
\left(\frac{d\psi_0}{dz}\right)_{z=(-a)^+}-\left(\frac{d\psi_0}{dz}\right)_{z=(-a)^-}=\frac{2m}{\hbar^2}g^{-}_{_{1D}}\psi_0(z=-a)
\,\,,
\end{eqnarray}
which are obtained by integrating Eq.(\ref{effective-1D-Schrodinger eq}) over $z$ and in the small range around $z=a$ and $z=-a$.

From the boundary condition (\ref{jump-condition-1},\ref{jump-condition-2}), it is obvious that if $g^{\pm}_{1D}$ tends to infinity, then $\psi_0(z=\pm a)$ must be equal to zero. So deverging $g^{\pm}_{1D}$ implies vanishing of the wave function at atom-impurity zero separation; hence, total atom-impurity reflection (zero transmission) would occur which is the indication of CIR.

\section{Results and Discussion} \label{Results}

From the $g^{+}_{1D}$ definition (Eq.\ref{g1D definition}) and the total transmission $T_{tot}=|1+f_e+f_o|^2$, it is clear that CIR occurs whenever $1+f_e+f_o=0$, or equivalently at complete reflectance $R=|f_e-f_o|^2=1$ where $g^{-}_{1D}$ diverges in the limit $k\rightarrow 0$ (Eq.\ref{g1D definition-2}).

Implying $f_e+f_o=-1$ leads us to a quadractic equation with respect to the ratio $a_{\perp}/a_{_{3D}}$.  So the CIR condition is defined as follows
\begin{eqnarray}\label{CIRposition-total}
\frac{a_{\perp}}{a_{_{3D}}}&=&-\frac{\alpha}{2}+\frac{2a}{a_{\perp}}(1-\epsilon)\nn\\&\pm &\sqrt{\frac{\beta^2}{4}-\frac{\beta}{ka_\perp}\sin(2ka)+\frac{4a^2}{a_{\perp}^2}(1+{\gamma}^2)-\frac{2a\beta}{a_{\perp}}\left(\gamma +\cos(2ka)\right)+\frac{4a\gamma}{ka_{\perp}^2}\left(\sin(2ka)+2ka\cos(2ka)\right)}\nn \,\,,\\
\end{eqnarray}
where $\alpha=\widetilde{\Lambda}(0,\epsilon)=\zeta(1/2,1+\epsilon)$ , $\beta=\widetilde{\Lambda}(\frac{4a}{a_\perp},\epsilon)=-\frac{a_\perp}{2a}+\sum_{n=1}^{\infty}\frac{e^{-\frac{4a\sqrt{n+\epsilon}}{a_\perp}}}{\sqrt{n+\epsilon}} $ , and $\gamma=\widetilde{F}(\frac{4a}{a_{\perp}},\epsilon)=-\frac{a_{\perp}^2}{8a^2}+\sum_{n=1}^{\infty}e^{-\frac{4a}{a_{\perp}}\sqrt{n+\epsilon}} $ .

\begin{figure}[H]
\centering\includegraphics[width=\textwidth]{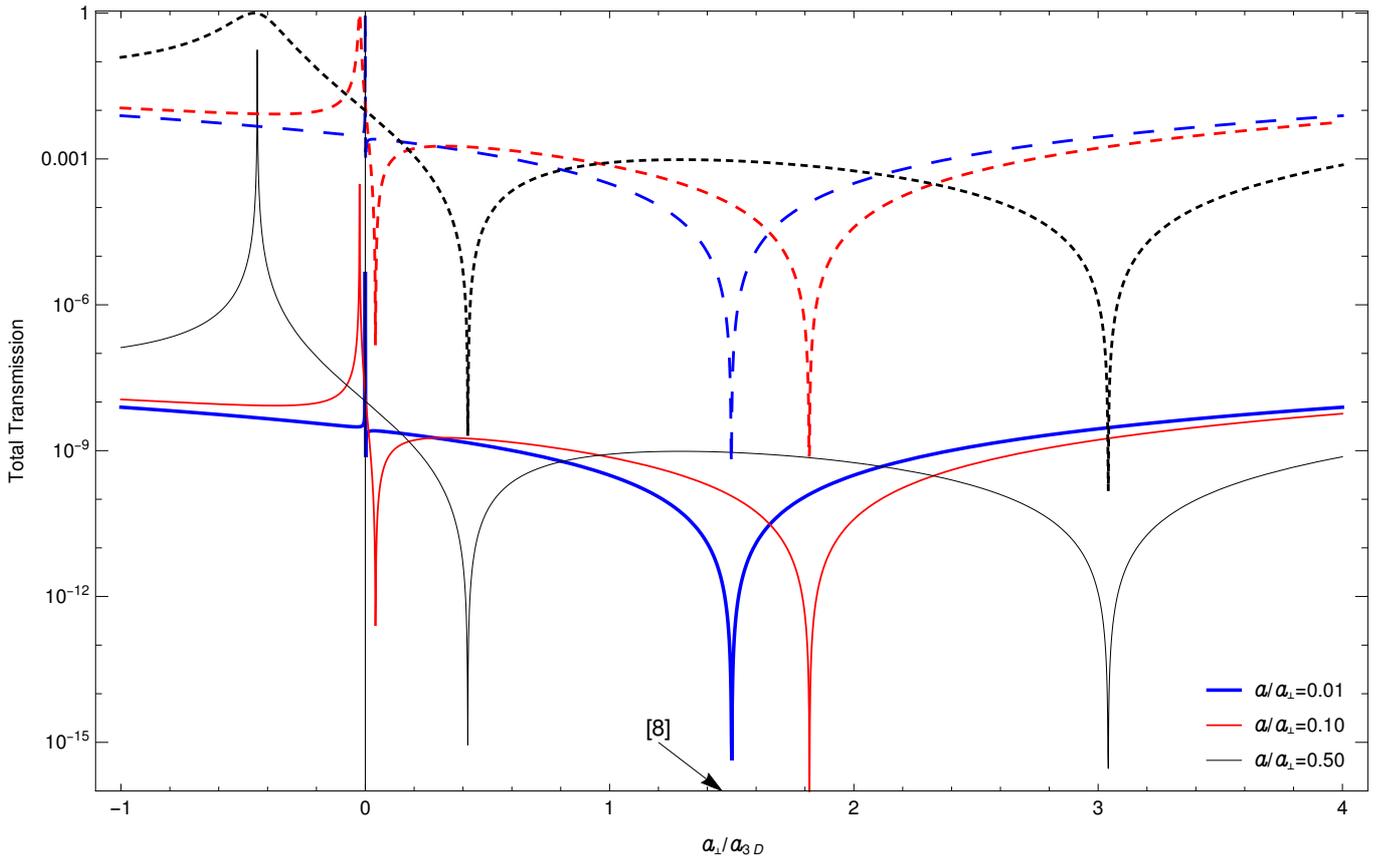}
\caption{ (Color online) Total transmission coefficient $T_{tot}(\frac{a_{\perp}}{a_{_{3D}}},a,k)$ (in logarithmic scale) for three different values of $a=0.01a_{\perp}$ (blue curves), $a=0.10a_{\perp}$ (red curves), and $a=0.50a_{\perp}$ (black curves). Dashed (solid) curves belong to $ka_{\perp}=0.0707$ ($ka_{\perp}=7.07\times 10^{-5}$). We indicate by arrow the CIR position obtained by Olshanii \cite{Olshanii} for the scattering in a one-center problem. }
\label{fig:Fig2}
\end{figure}

\begin{figure}[H]
\begin{center}
\begin{subfigure}{
\includegraphics[width=0.8\columnwidth]{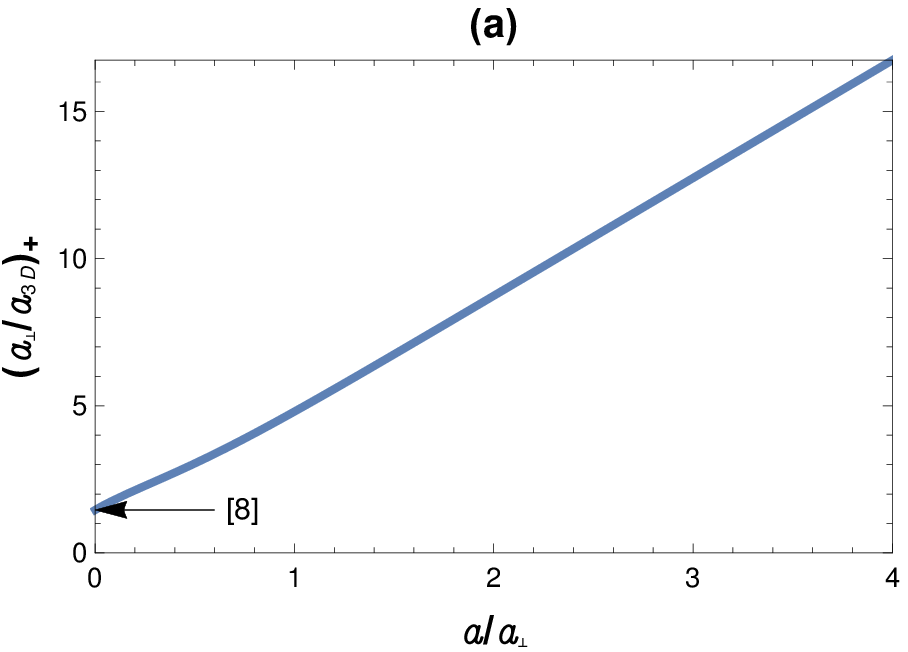}}
\end{subfigure}\\
\begin{subfigure}{
\includegraphics[width=0.8\columnwidth]{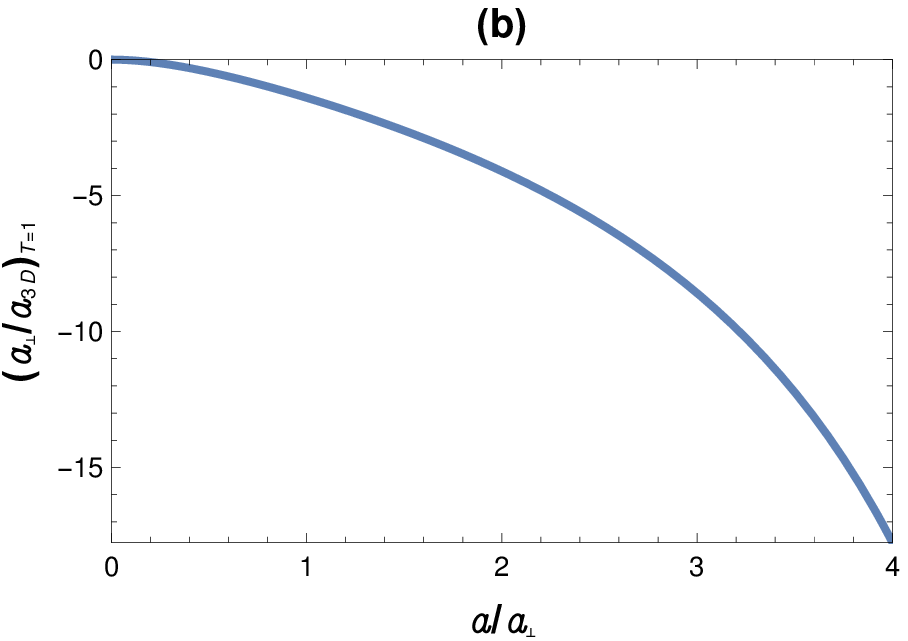}}
\end{subfigure}
\end{center}
\caption{ CIR position versus $a/a_{\perp}$ when both even and odd wave scattering are included and $ka_{\perp}=0.0707$. (a) The CIR condition when $T_{tot}=0 $ (obtained from Eq.(\ref{CIRposition-total}) with plus sign before the square root).The CIR position obtained by Olshanii \cite{Olshanii} for the scattering in a one-center problem is indicated with an arrow. (b) Dual CIR position when $T_{tot}=1 $.}
\label{fig:Fig3}
\end{figure}

In \figref{Fig2}, the graph for the total transmission $T_{tot}(\frac{a_{\perp}}{a_{_{3D}}},a,k)$ is shown as a function of the ratio $a_{\perp}/a_{_{3D}}$ for three distinct values $a/{a_{\perp}}=0.01,~0.1,~0.5$ and two $k$, $k=0.0707/a_{\perp}$ and $k=7.07\times 10^{-5}/a_{\perp}$. Every calculated curve $T_{tot}(\frac{a_{\perp}}{a_{_{3D}}},a,k)$ has two minima for each momentum $k$ and a fixed separation $2a$ between the scattering centers. The position of the minima exactly coincides with the values obtained form Eq.(\ref{CIRposition-total}) which by definition correspond to the positions of CIRs. The right and left minima of the graph correspond to the values with plus and minus sign before the square root in Eq.(\ref{CIRposition-total}), respectively. Thus, the position ${(a_{\perp}/a_{_{3D}})}_+$ , corresponding to the right minimum in $T_{tot}(\frac{a_{\perp}}{a_{_{3D}}},a,k)$, demonstrates rather strong dependence on the separation $2a$ between the scattering centers (see \figref{Fig3}(a)) and in the limit $a\rightarrow 0$ approaches the value $a_{\perp}/a_{_{3D}}=1.46$ which was obtained by Olshanii \cite{Olshanii} for the CIR position in the even state for the case of single scatterer. This fact leads to a conclusion that the right minimum in the total transmission originates in the even state scattering. As can be seen in \figref{Fig2}, the position of minima and maxima in the transmission curve for the range $a_{\perp}k\leqslant 0.0707$ is slightly dependent on the $k$ value.

\begin{figure}[H]
\begin{center}
\begin{subfigure}{
\includegraphics[width=0.8\columnwidth]{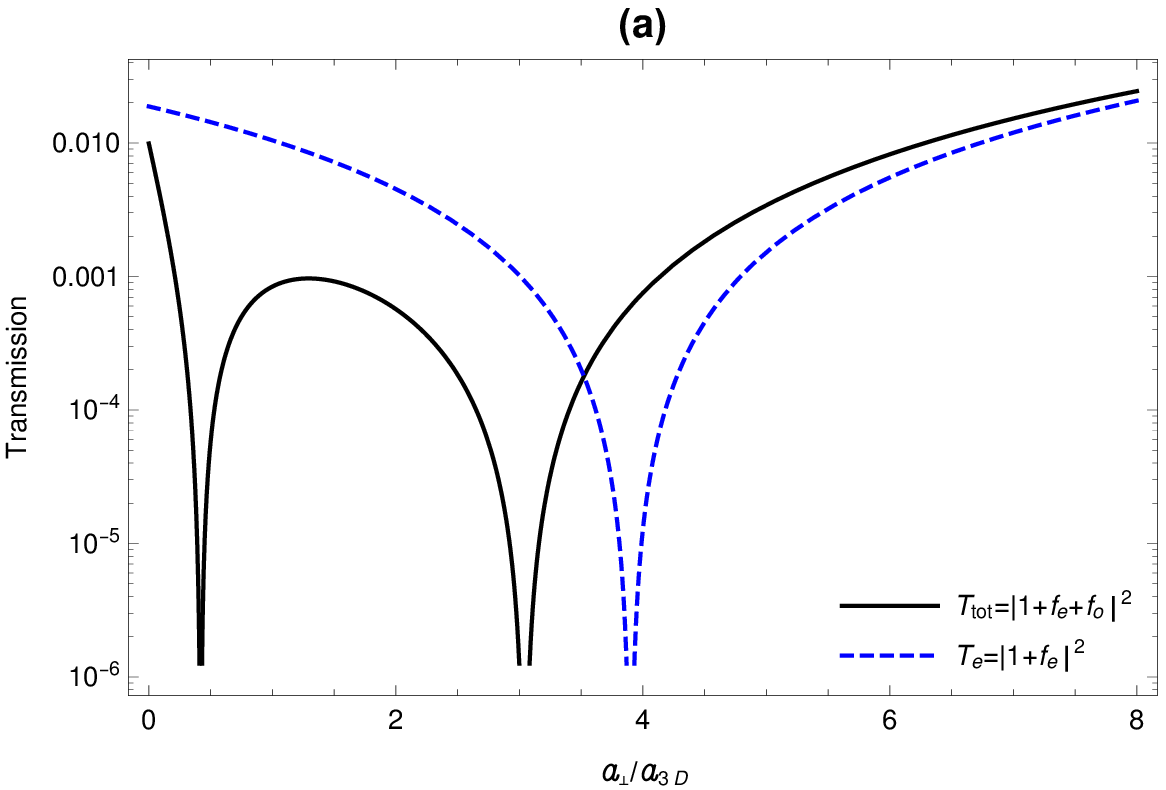}}
\end{subfigure}\\
\begin{subfigure}{
\includegraphics[width=0.8\columnwidth]{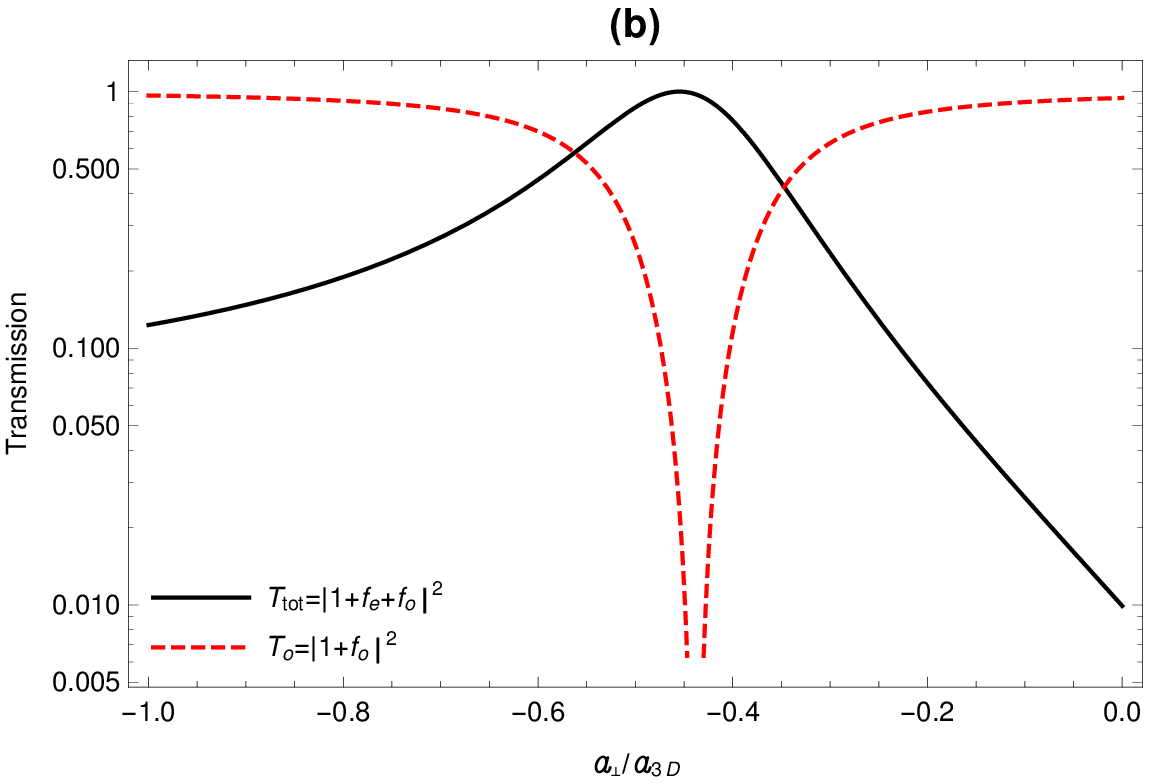}}
\end{subfigure}
\end{center}
\caption{ (Color online) Total transmission coefficient $T_{tot}$ (in logarithmic scale) along with (a) pure even transmission $T_e$ (b) pure odd transmission $T_o$ for the impurity separation $2a=a_{\perp}$ as a function of $a_{\perp}/a_{_{3D}}$ when $ka_{\perp}=0.0707$  .}
\label{fig:Fig4}
\end{figure}

\begin{figure}[H]
\centering\includegraphics[width=0.8\textwidth]{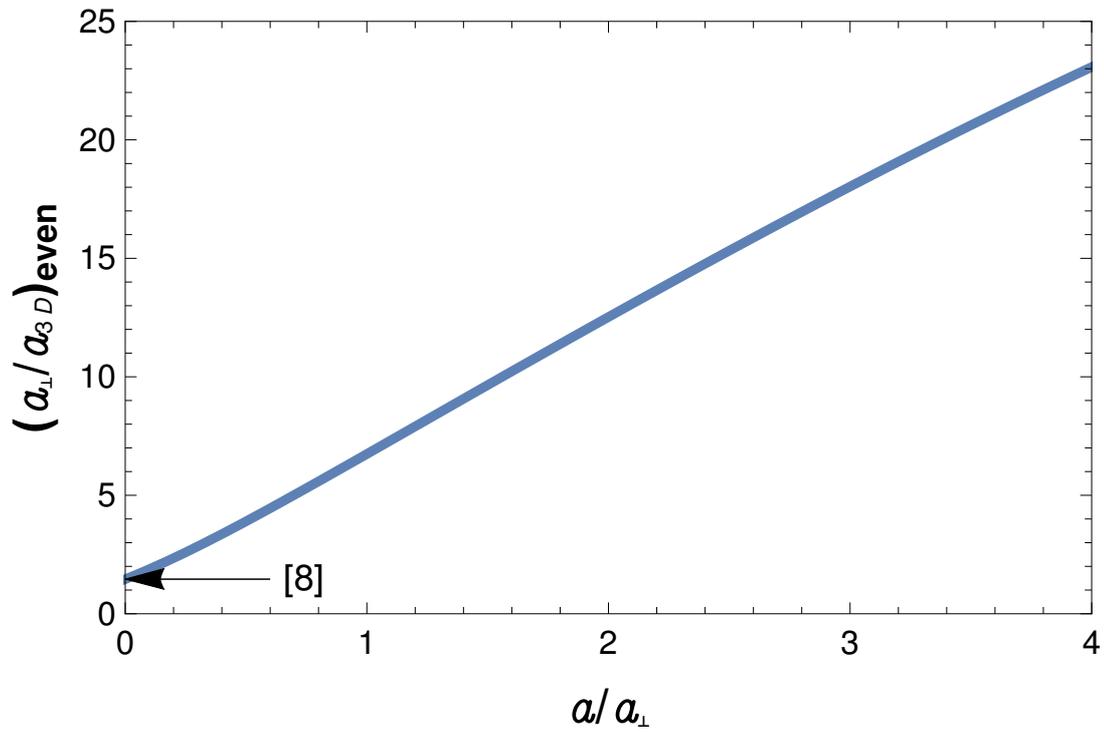}
\caption{ Even CIR position versus $a/a_{\perp}$ when $ka_{\perp}=0.0707$. The CIR position obtained by Olshanii \cite{Olshanii} for the scattering in a one-center problem is here indicated with an arrow. }
\label{fig:Fig5}
\end{figure}

To clarify the origin of the left minimum and the maximum in the total transmission curve, we have also analyzed the partial transmissions $T_e=|1+f_e|^2$ and $T_o=|1+f_o|^2$. In \figref{Fig4}, the total transmission curve is plotted along with the transmission in the even $T_e=|1+f_e|^2$ and odd $T_o=|1+f_o|^2$ states for $a=0.5a_{\perp}$. From \figref{Fig4}(a) it is clear that the right minimum of $T_{tot}$ is close to the minimum of $T_e$ but is shifted. It means that the even part of the scattering wave gives the main contribution to the scattering and the odd part of the scattering wave is almost negligible in that specific range of the ratio $a_{\perp}/a_{_{3D}}$ \cite{Melezhik2007}. For pure even scattering, the CIR occurrence requires $T_e=|1+f_e|^2=0$ , yielding the following position of the ``even" CIR
\begin{eqnarray}
\left(\frac{a_{\perp}}{a_{_{3D}}}\right)_e=-\frac{1}{2}(\alpha +\beta)+\frac{\sin(2ka)}{ka_{\perp}}+\frac{4a}{a_{\perp}}{\cos}^2(ka)+\frac{2a}{a_{\perp}}(\gamma -\epsilon) \,\,.
\end{eqnarray}
The dependence of the ``even" CIR  position on the separation of impurities is demonstrated in \figref{Fig5}. It qualitatively repeats the behaviour of the right CIR in the total transmission shown in \figref{Fig3}(a).

\begin{figure}[H]
\centering\includegraphics[width=0.8\textwidth]{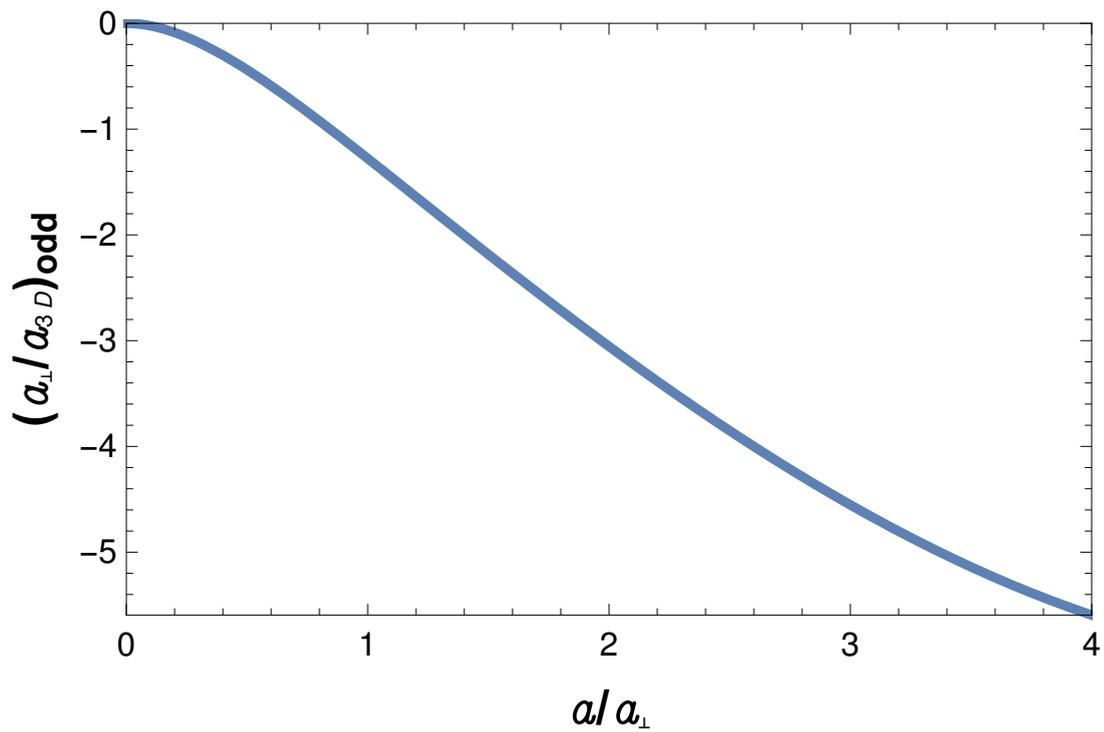}
\caption{ Odd CIR position versus $a/a_{\perp}$ when $ka_{\perp}=0.0707$. }
\label{fig:Fig6}
\end{figure}

On the other hand, the minimum of odd transmission $T_{o}$ coincides with the position of complete total transmission (\figref{Fig4}(b)). This phenomenon occurs as a result of destructive interference between odd and even scattering waves in the waveguide trap yielding a zero reflectance ($R=1-T=0$), the so-called dual CIR \cite{Kim2006, Kim2007}. In our case, $R=|f_e-f_o|=0$ leads to an equation for the dual CIR condition
\begin{eqnarray}\label{dual CIR position}
{(\frac{a_{\perp}}{a_{_{3D}}})}_{T=1}=\frac{1}{\cos(2ka)-2ka\sin(2ka)}\Bigg[(4k\frac{a^2}{a_{\perp}}(\epsilon -1)+ka\alpha)\sin(2ka)\nn\\ \nn\\
-\frac{\sin(2ka)}{ka_{\perp}}-\frac{1}{2}(\alpha\cos(2ka)-\beta)-\frac{2a}{a_{\perp}}(\epsilon\cos(2ka)+\gamma) \Bigg] \,\,,
\end{eqnarray}
illustrated in \figref{Fig3}(b) and \figref{Fig4}(b), and the position of the odd CIR ($T_o=|1+f_o|^2=0$) obeys the following relation:
\begin{eqnarray}
\left(\frac{a_{\perp}}{a_{_{3D}}}\right)_o=-\frac{1}{2}(\alpha -\beta)-\frac{\sin(2ka)}{ka_{\perp}}+\frac{4a}{a_{\perp}}{\sin}^2(ka)-\frac{2a}{a_{\perp}}(\gamma +\epsilon) \,\,,
\end{eqnarray}
which is shown in \figref{Fig6} with a quite similar behaviour to the dual CIR position in \figref{Fig3}(b).

\begin{figure}[H]
\begin{center}
\begin{subfigure}{
\includegraphics[width=0.8\columnwidth]{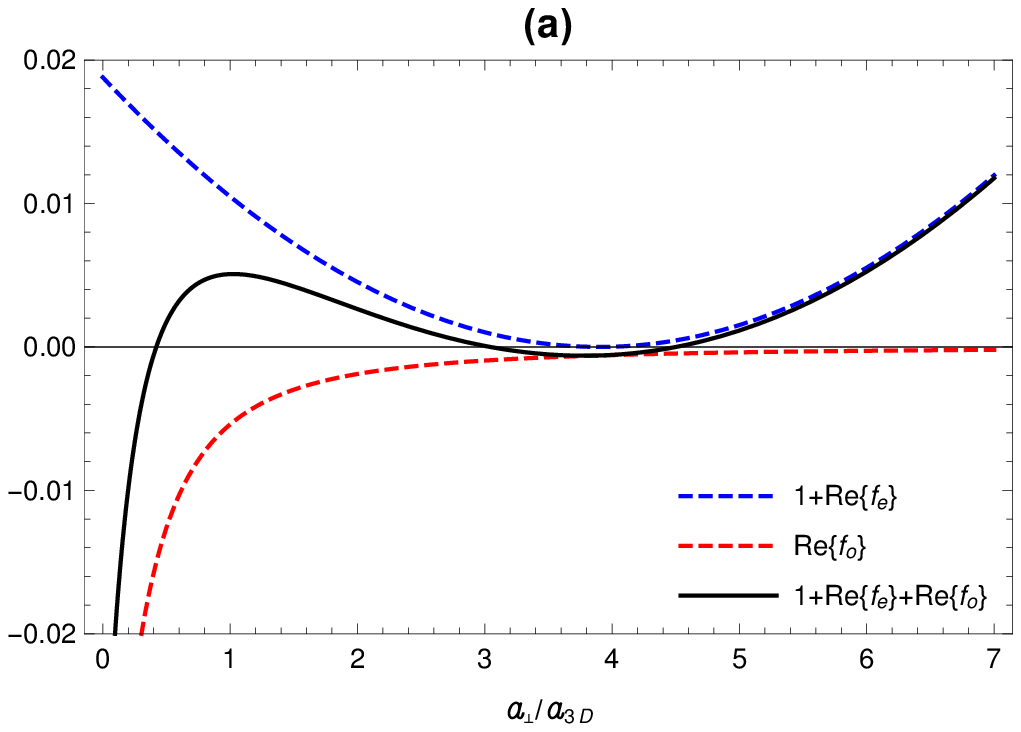}}
\end{subfigure}\\
\begin{subfigure}{
\includegraphics[width=0.8\columnwidth]{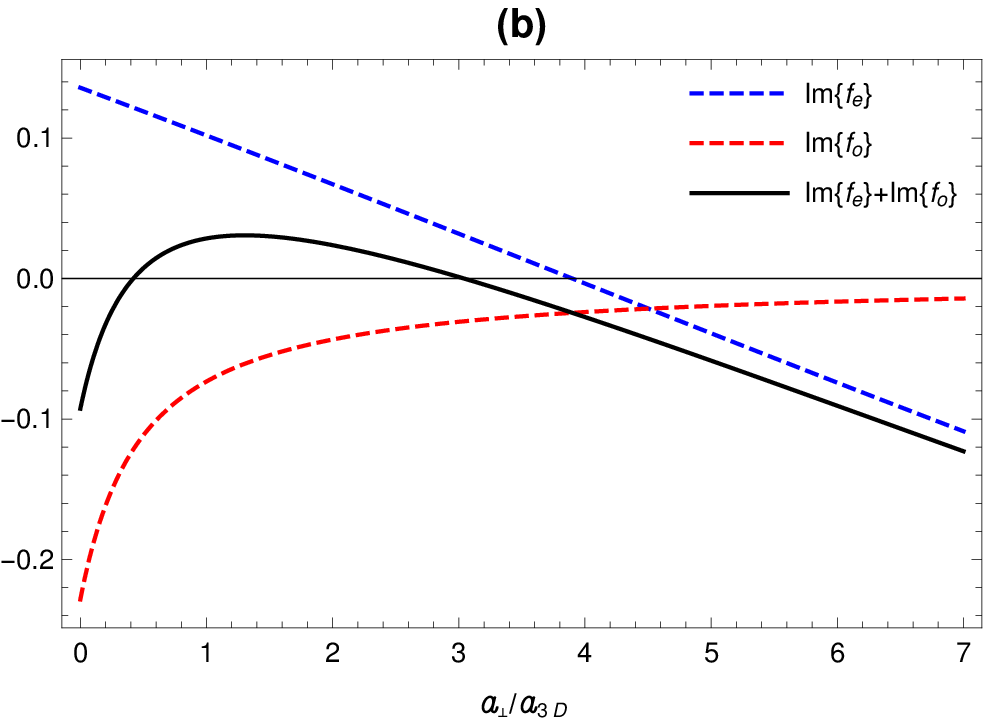}}
\end{subfigure}
\end{center}
\caption{ (Color online) (a) $\left[1+Re\{f_e\}\right]$ (blue dashed curve), $Re\{f_o\}$ (red dashed curve) and the resultant $\left[1+Re\{f_e\}+Re\{f_o\}\right]$ (black solid curve) (b) $Im\{f_e\}$ (blue dashed curve), $Im\{f_o\}$ (red dashed curve) and the resultant $\left[Im\{f_e\}+Im\{f_o\}\right]$ (black solid curve) for the impurity separation $2a=a_{\perp}$ as a function of $a_{\perp}/a_{_{3D}}$ when $ka_{\perp}=0.0707$. Both black solid curves pass zero axis at $a_{\perp}/a_{_{3D}}=0.42,~3.04$, representing the CIR positions.}
\label{fig:Fig7}
\end{figure}

We have to note that the presence of odd scattering is also ``responsible" for the left minimum in the total transmission $T_{tot}$ in the region of $a_{\perp}/a_{_{3D}}$ where an interplay between odd and even scattering amplitudes results in $\left[1+f_e+f_o\right]=0$ and subsequently in zero transmission (see \figref{Fig7}).

With decreasing distance between the scatterers (as $a\rightarrow 0$), the maximum and left minimum in the total transmission approach one another (see \figref{Fig2}). However, our computational scheme is not applicable to the odd scattering at $a=0$ \cite{Blume}. Nevertheless, in the vicinity of $a=0$ (as $a\rightarrow 0$) our model gives a qualitatively correct result for the odd scattering. To solve the problem quantitatively as $a\rightarrow 0$ for the case where the particle-impurity interaction range is of considerable amount, one has to take into account the actual shape of the interaction potential.

\section{Conclusion}\label{Conclusion}
We have investigated atomic scattering from two centers (impurities) fixed on the longitudinal axis of a waveguide-like trap via the pseudopotential approach. In this method, we have introduced a new regularization operator for the zero-range interaction potential leading to consistent results with the corresponding one-center problem when the displacement between the scattering centers approaches zero.

We have found that in contrast to the confined scattering on a single center in the s-wave pseudopotential approach \cite{Olshanii}, there are two CIRs in the confined two-center problem due to the resonances in the even and odd scattering states at a fixed distance between the centers. Moreover, with increasing the distance, the CIR position is shifted to higher values of the ratio $a_{\perp}/a_{_{3D}}$.

The obtained results can be considered as a starting point for quantitative analysis of confined atomic scattering on fixed impurities or two-atomic molecules. The analysis can be improved by using more realistic potentials for atom-impurity (atom-molecule) interactions.

The regularization method we suggested here can be extended to the case of $N$ impurities in a waveguide-like trap, thus paving a way to solve many-body problems in confined geometries where interactions in the corresponding nonlinear Schr\"odinger equation can be simulated by pseudopotentials. It can also be useful for constructing a mean-field approach without ultraviolet divergencies due to contact interactions \cite{Olshani-Pri2001}.

\section{Acknowledgements}
We thank Z. Idziaszek, V. I. Korobov, A. Negretti, V. V. Pupyshev, P. Schmelcher, S. Yu. Slavyanov, and F. Vukajlovic for valuble discussions. This work was supported by the Russian Foundation for Basic Research, Grant No. 18-02-00673 and the ``RUDN University Program 5-100''.
\section{Appendix A}
For simplifying the second derivatives appearing in the definition of pseudopotential (Eq.(\ref{V_3D_elaborated})), we proceed as follows:\\
By using
\begin{eqnarray}
\frac{\partial z}{\partial r_1}&=&\frac{z-a}{r_1}\\
\frac{\partial \rho}{\partial r_1}&=&\frac{\rho}{r_1}
\end{eqnarray}
and
\begin{eqnarray}
\frac{\partial z}{\partial r_2}&=&\frac{z+a}{r_2}\\
\frac{\partial \rho}{\partial r_2}&=&\frac{\rho}{r_2}
\end{eqnarray}
we get
\begin{eqnarray}
\frac{\partial}{\partial r_1}=\frac{\rho}{r_1}\frac{\partial}{\partial\rho}+\frac{z-a}{r_1}\frac{\partial}{\partial z}\\
\frac{\partial}{\partial r_2}=\frac{\rho}{r_2}\frac{\partial}{\partial\rho}+\frac{z+a}{r_2}\frac{\partial}{\partial z}
\end{eqnarray}
Since $\eta_1$ and $\eta_2$ are defined in the limit $r_{1(2)}\rightarrow 0$ (29,30), the derivatives should be computed at $\rho=0$, so we deal with
\begin{eqnarray}\label{partial-derivative}
\frac{\partial}{\partial r_1}=\frac{z-a}{|z-a|}\frac{\partial}{\partial z}\\
\frac{\partial}{\partial r_2}=\frac{z+a}{|z+a|}\frac{\partial}{\partial z}
\end{eqnarray}
In the limit $r_1\rightarrow 0$ and $r_2\rightarrow 0$, we will have $z\rightarrow a$ and $z\rightarrow -a$ , respectively:
\begin{eqnarray}
\eta_1=\frac{1}{2}\frac{\partial}{\partial{r_1}}\left[\frac{\partial}{\partial{r_2}}(r_1 r_2\psi) \right]_{r_1\rightarrow 0}&=&\frac{1}{2}\frac{z-a}{|z-a|}\frac{\partial}{\partial z}\left[\frac{z+a}{|z+a|}\frac{\partial}{\partial z}\left(|z-a||z+a|\psi(\rho=0,z) \right)\right]_{z\rightarrow a}\nn\\&=&\frac{1}{2}\frac{d^{2}}{dz^2}\left[(z-a)(z+a)\psi(\rho=0,z) \right]_{z\rightarrow a^+}
\end{eqnarray}
\begin{eqnarray}
\eta_2=\frac{1}{2}\frac{\partial}{\partial{r_2}}\left[\frac{\partial}{\partial{r_1}}(r_2 r_1\psi) \right]_{r_2\rightarrow 0}&=&\frac{1}{2}\frac{z+a}{|z+a|}\frac{\partial}{\partial z}\left[\frac{z-a}{|z-a|}\frac{\partial}{\partial z}\left(|z-a||z+a|\psi(\rho=0,z) \right)\right]_{z\rightarrow {-a}^-}\nn\\&=&\frac{1}{2}\frac{d^2}{dz^2}\left[(z-a)(z+a)\psi(\rho=0,z) \right]_{z\rightarrow {-a}^-}
\end{eqnarray}


\end{document}